\documentstyle[prd,aps,epsf]{revtex}
\begin{document}

\title{Gravitational instantons, extra dimensions and form fields.}

\author{James Gray$^1$ and E. J. Copeland$^1$} 
\address{$^1$Centre for Theoretical Physics, University of Sussex, Brighton, 
BN1 9QH, Great Britain}  

\date{\today}  
\maketitle 

\begin{abstract}
A broad class of higher dimensional instanton solutions are found for a theory which contains gravity, a scalar field 
and antisymmetric tensor fields of arbitrary rank. The metric used, a warp product of an arbitrary number of any compact 
Einstein manifolds, includes many of great interest in particle physics and cosmology. For example 4D FRW universes with 
additional dimensions compactified on a Calabi-Yau three fold, a torus , a compact hyperbolic manifold or a sphere are 
all included. It is shown that the solution of this form which dominates the Hartle Hawking path integral is always a 
higher dimensional generalisation of a Hawking Turok instanton when the potential of the scalar field is such that these 
instantons can exist. On continuation to lorentzian signature such instantons give rise to a spacetime in which all of 
the spatial dimensions are of equal size and where the spatial topology is that of a sphere. The extra dimensions are 
thus not hidden. In the case where the potential for the scalar field is generated solely by a dilatonic coupling to the 
form fields we find no integrable instantons at all. In particular we find no integrable solutions of the type under 
consideration for the supergravity theories which are the low energy effective field theories of superstrings.
\end{abstract} 

\section{Introduction.}

Recently a number of papers have appeared in which the instanton cosmology associated with higher dimensional theories 
is considered \cite{hawking98,stelle2,garriga98,chiba,stelle1,HR,us,wesson}. Interest in such a union of initial 
conditions and modern particle physics theories is understandable. Extra hidden dimensions are a vital ingredient of 
many modern models of particle physics yet there are a number of ways in which a higher dimensional theory can be 
dimensionally reduced, leading to many possible effective theories in 4D at low energies. This loss in predictive power 
could be reversed if a well motivated proposal for the state of the universe existed which picked out a particular 
compactification scheme.

One such proposal that has received considerable attention is due to Hartle and Hawking \cite{hartle}. Recently Hawking 
and Turok \cite{hawking98,hawking98a} obtained a new class of instanton solutions which, using this proposal, could be 
interpreted as representing the initial configurations for our universe. In an earlier work with Saffin \cite{us} we 
addressed the issue of the singular instantons arising in higher dimensional theories, and showed that in the presence 
of only a scalar field, the most likely configuration we found was always the highly symmetrical Hawking Turok instanton 
- a possible problem when we wish to have compactified extra spatial dimensions.

In this paper we address the consequences of adding extra matter fields, namely form fields, into the higher dimensional 
action. The calculations peformed here and in \cite{us} and \cite{saf} can be viewed as a necessary complement to the work 
which 
addresses the important issue of perturbations about instantons 
\cite{gratton,hertog,hertog2,gratton2,gratton3,tanaka,george}. Such calculations can show whether or not a given 
solution is a local minimum of the action. However, there may be other minima of lower action elsewhere in parameter 
space where the minima under consideration are in different conformal equivalence classes. It would then be one or more 
of these other minima which would dominate the path integral, not the first solution.

The layout of the paper is as follows. In section II, we introduce our metric and briefly recap the results for the case 
where the form fields are absent \cite{us}. In section III we proceed to introduce antisymmetric fields through a  
lagrangian which is a simple field strength squared term. In section IV we examine the case of more general 
couplings between the scalar and antisymmetric tensor fields. We also make some comments about the absence of integrable 
instantons of the type under consideration in the low energy effective field theories of string theory before conluding 
in section V.

\section{Instantons without form fields.}

Our starting point is a manifold ${\cal M}$ which has a metric
structure imposed on it, and a scalar field $\phi$ living on it
with potential ${\cal V}(\phi)$.
By using the usual torsion free, metric connection on ${\cal M}$ we are able to
describe the equations of motion for both the metric and $\phi$,
which follow from the Einstein-Hilbert action.
\begin{eqnarray}
\label{EHaction} S_{\rm{E}}&=&\int_{\cal M} \eta
\left[-\frac{1}{2\kappa^2}{\cal R}+\frac{1}{2}(\partial \phi)^2 +{\cal
V}(\phi)\right]+\textnormal{boundary term}.
\end{eqnarray}
Here, $\kappa^2 = 8\pi /m_{\rm Pl}^2$ (scaled to
unity for the rest of  the paper), and the
boundary terms are such that the action does not
contain second derivatives of the
metric\cite{gibbons77}\cite{barrow89}. $\eta$ is the volume form
and ${\cal R}$ is the Ricci scalar of the connection.

We consider the manifold
${\cal M}$ as a foliation in
Euclidean time of a product of boundary-less manifolds. At any
given time $\xi$ we can then write ${\cal M}(\xi)$ as a 
Cartesian product of
${\cal M}_{(i)}$, $i=1...T$ each with dimension $n_{(i)}$, where
for convenience we define \mbox{$N=n_{(1)}+n_{(2)}+...+n_{(T)}$}.
To endow ${\cal M}$ with a metric structure we start by putting a
metric on each of the ${\cal M}_{(i)}$, denoted $\rm{ds}_{(i)}^2$.
The metric structure we impose on ${\cal M}$ then follows by
introducing a $\xi$ dependent scale factor, $a_{(i)}$,
for each ${\cal M}_{(i)}$;
providing information about the relative size of the
${\cal M}_{(i)}$ at any given $\xi$. The resulting metric is then,

\begin{eqnarray}
\label{M_metric}
\rm{ds}^2&=&\rm{d}\xi^2+a_{(1)}^2(\xi)\rm{ds}_{(1)}^2
                       +a_{(2)}^2(\xi)\rm{ds}_{(2)}^2
                       +...+a_{(T)}^2(\xi)\rm{ds}_{(T)}^2.
\end{eqnarray}

The only restriction on the $ds_{(i)}^2$'s is that they
are Einstein metrics on compact manifolds; the Ricci tensor 
is proportional to the
metric, the constant of proportionality being $\Lambda_i$ for the $i^{th}$ submanifold. This is necessary if the metric 
is to be consistent with 
our ansatz for the scalar field, $\phi = \phi(\xi)$. Of the many solutions that 
exist, a class of these instantons can be continued to
a four dimensional inflating universe, with a number of static extra dimensions.
This form for the metric is very general. It includes a wide class
of metrics commonly considered in cosmology, such as those leading
to the Coleman-De Lucia instanton \cite{coleman80},
Kantowski-Sachs instantons \cite{jensen89}, Hawking-Turok
instanton \cite{hawking98}, most of the models of string
cosmology arising out of compactifications on tori (for a review
see \cite{lidsey99}), compactifications of string theory on
Calabi-Yau spaces (for a review see \cite{polchinski}), and some
compactifications of Supergravity theories on spheres (for a review
see \cite{duff}). For example in \cite{hawking98}
\cite{coleman80} $T=1$ and ${\cal M}_{(1)}$ is a
three sphere with its standard round metric. A more exotic
Kantowski-Sachs metric was considered in \cite{jensen89}, there
$T=2$ with ${\cal M}_{(1)}=S^1$, ${\cal M}_{(2)}=S^2$.

The equations of motion for this system are as follows (see \cite{us} for details).

\begin{equation}
\label{ab_constraint8}
\frac{1}{2} \left( \sum_{i=1}^T n_i \frac{a'_i}{a_i} \right)^2 - \frac{1}{2} 
\sum_{i=1}^T n_i \left( \frac{a'_i}{a_i} \right)^2 - \sum_{i=1}^T \frac{n_i 
\Lambda_i}{2 a_i^2} = \frac{1}{2} \phi'^2 - {\cal V}
\end{equation}
\begin{eqnarray}
\label{aeqn8}
\sum_{i=1}^T n_i \frac{a''_i}{a_i} - \frac{a''_k}{a_k} + \frac{1}{2} \left( 
\sum_{i=1}^T n_i \frac{a'_i}{a_i} \right)^2 - \sum_{i=1}^T n_i \frac{a'_i 
a'_k}{a_i a_k} - \frac{1}{2} \sum_{i=1}^T n_i \left( \frac{a'_i}{a_i} \right)^2 
+ \left( \frac{a'_k}{a_k} \right)^2 - \frac{1}{2} \sum_{i=1}^T \frac{n_i 
\Lambda_i}{a_i^2} + \frac{\Lambda_k}{a_k^2} \\
\nonumber = - \frac{1}{2} \phi'^2 - {\cal V} 
\end{eqnarray}
\begin{equation}
\label{phieqn8}
\phi''+ \sum_{i=1}^T n_i \frac{a_i'}{a_i} \phi' = \frac{\partial {\cal
V}}{\partial \phi},
\end{equation}

where Eq. (\ref{aeqn8}) is true for all k. By defining $V_{(i)}$ as the volume of ${\cal M}_{(i)}$ and
$\beta$ to be the product of the scale factors we find
that the action, including boundary terms, is given by
\begin{eqnarray}
\label{tot_action9}
S_{\rm{E}}&=&V_{(1)}V_{(2)}...V_{(T)}\left\{
-\left[\frac{\partial\beta}{\partial \xi}\right]_{\xi_S}^{\xi_{N}}
+\int {\rm d}\xi\beta\left[-\frac{1}{2}{\cal R}
                     +\frac{1}{2}\phi'^2+{\cal V}(\phi)\right]
\right\}\\
\label{eff_action9}
&=&V_{(1)}V_{(2)}...V_{(T)}\left\{
-\left[\frac{\partial\beta}{\partial \xi}\right]_{\xi_S}^{\xi_{N}}
-\frac{2}{n_{(1)}+n_{(2)}+...+n_{(T)}-1}\int {\rm d}\xi \beta(\xi){\cal V(\phi)}
\right\}\\
\label{beta}
\beta(\xi)&=&a_{(1)}^{n_{(1)}}a_{(2)}^{n_{(2)}}...a_{(T)}^{n_{(T)}}.
\end{eqnarray}
In arriving at the second equation we have used the trace of
Einstein's equations to eliminate 
the scalar curvature and scalar field kinetic terms. The quantities
$\xi_S$ and $\xi_N$ refer to the range of the $\xi$ coordinate, with
$\xi_N$ being the `north' pole and $\xi_S$ referring to the `south' 
pole of the instanton taken to be $\xi=0$. To save writing out the
volumes of all the submanifolds we shall call the term in the curly
braces of Eq. (\ref{eff_action9}) the {\it reduced action}.

Given that the metrics on the ${\cal M}_i$'s are Einstein, the evolution equations for the scale
factors $a_{(i)}(\xi)$ depend only on the value of the
`cosmological constants' $\Lambda_{(i)}$ and not on the detailed
topology or geometry of the manifold. 
This is potentially very significant,
for any statements we can make about the evolution of the scale
factors cover a very large class of manifolds, all those admitting
an Einstein metric.

In \cite{us} it was concluded that the instanton of the above class with 
minimum Euclidean action is a higher dimensional generalisation of the Hawking 
Turok solution. It was also conjectured that the action of a configuration could only be negative 
if all of its submanifolds were positively curved. The reason for this conjecture, which we checked for a wide range of 
solutions, is that negative or 
zero curvature submanifolds can have arbitrary volume for a given curvature. 
Therefore, since the action is proportional to the volume of each of the 
submanifolds if we had an instanton with negative action and a negatively curved 
submanifold we could make its action arbitrarily negative by raising the volume 
of that submanifold. Clearly this would make our instanton approximation to the 
path integral ill defined. In the case of positively curved sub manifolds this 
problem does not occur. Bishop's theorem states that for a positively curved 
Einstein manifold of given curvature the metric which maximises the volume of 
the manifold is the round sphere. Thus there is an upper limit to the volume of 
the submanifolds, which is saturated by the Hawking Turok type of solutions.

The conclusion that a Hawking Turok instanton dominates the path integral in 
these higher dimensional theories leads to a potential problem. When this instanton is 
continued to Lorentzian signature it gives a spacetime which is a higher 
dimensional generalisation of an open FRW. Thus the Hartle Hawking proposal in 
the case where we have this simple matter content predicts that none of the 
extra dimensions should be compactified.

While this metric is very general the matter field content, just a minimally 
coupled scalar field, is not.
The obvious thing to study then is whether or not this problem arises in situations where 
more general forms of 
matter are present. Preliminary calculations for the cases of multiple and non-minimally coupled scalar fields showed 
that such simple modifications of the theory considered in \cite{us} are not enough to solve this problem. We therefore 
turn to another common ingredient of the bosonic sectors of higher dimensional particle physics theories - antisymmetric 
tensor fields - will the extra structure they introduce give us more phenomenologically acceptable results?

\section{Instantons with form fields.}

	Antisymmetric tensor fields occur commonly in 
the bosonic sectors of supergravity theories and can be used 
to obtain solutions where some of the dimensions of the theories are 
compactified \cite{Freund}. These fields have also been used in a cosmological setting where the dominant instanton is 
assumed to be one which gives rise to a compactified spacetime \cite{hawking98,stelle1,HR}.

We shall start therefore by considering the following d-dimensional action (temporarily reinserting $\kappa$).

\begin{eqnarray}
\label{EHFaction} S_{\rm{E}}&=&\int_{\cal M} d^d x \sqrt{g} \left[-\frac{1}{2\kappa^2}{\cal R}+\frac{1}{2}(\partial 
\phi)^2 +{\cal V}(\phi) +\frac{1}{2 (d_f)!} F_{\mu_1...\mu_{d_f}} F^{\mu_1...\mu_{d_f}} \right]+\textnormal{GH + FF 
boundary terms}
\end{eqnarray}

where the GH (Gibbons Hawking) boundary term is given by,

\begin{eqnarray}
\textnormal{GH boundary term} = \frac{1}{\kappa^2} \int_{\cal \partial M} d^{d-1} x \sqrt{h} K
\end{eqnarray}

and the FF (Form Field) boundary term is given by,

\begin{eqnarray}
\textnormal{FF boundary term} = -\int_{\cal \partial M} d^{d-1} x \sqrt{h} 
\frac{1}{d_f!} F^{\mu_1...\mu_{d_f}} n_{\mu_1} A_{\mu_2...\mu_{d_f}}
\end{eqnarray}
 $A_{\mu_2...\mu_{d_f}}$ here is the antisymmetric potential which gives rise to the field strength F 
and n is a unit vector normal to the boundary. $K = h^{ij} K_{ij}$ is the trace 
of the second fundamental form, the trace being taken with the metric, $h_{ij}$, which has been induced on 
the boundary by $g_{ij}$.
 
 The boundary terms are vital. They ensure that the action we are considering leads to the equations of motion of the 
fields upon variation in a volume when we hold the metric and field strength, but not the derivative of the metric or 
the gauge potential, constant on the boundary. In other words the boundary terms ensure that the equations of motion in 
the classical limit follow in the usual way from the path integral \cite{gibbons77,ross}.
 
 The dimensionality of the antisymmetric tensor field, $d_f$, is important in 
determining which subset of the metrics seen in the previous section are 
consistent with supporting a non-zero field. There are two different Ansatze for the form fields which are consistent with 
the symmetries of our metric. They are \cite{low},

\begin{eqnarray}
\label{anz1}
F_{0 \mu_1.... \mu_{d_f-1}} = \frac{1}{\prod_{i=\textnormal{lit}}a_i^{n_i}} 
E'(\xi) 
\epsilon_{\mu_1.... \mu_{d_f-1}}
\end{eqnarray}

\begin{eqnarray}
\label{anz2}
F_{\mu_1... \mu_{d_f}} = \frac{1}{\prod_{i=\textnormal{lit}} a_i^{n_i}} M 
\epsilon_{\mu_1.... 
\mu_{d_f}}
\end{eqnarray}

The $i=$ lit in the products of scalefactors in the denominators simply mean that we only 
include the scale factors of manifolds 
which are spanned by the forms. \footnote{ A submanifold is spanned by a form if the form is non-zero when some or all 
of its indices correspond to coordinates associated with that submanifold.} $\epsilon^{\mu_1.... 
\mu_{d_f}}$ is a totally antisymmetric quantity the components of which vanish if any of the indices are not on one of 
the submanifolds spanned by 
the form and are $\pm 1$ otherwise. M is a constant and $E'$ is a function of $\xi$. For the sake of simplicity all the 
results we will present here have only one such ansatz activated per form in any given instanton. The more general case 
will be the subject of future work.

The first of these two ansatze is the generalisation of an electric field while 
the second is the generalisation of a magnetic field. We will refer to them as 
the electric and magnetic cases respectively from now on.

There are certain rules for combining forms of a given degree with submanifolds 
of a given dimensionality. These rules come from demanding that our form field 
ansatz combines with our metric (and scalar field) ansatz in such a way as to 
be a solution to Einstein's equations. Given the symmetries of our metric we 
have the following conditions.

\begin{itemize}
\item If we have a magnetic ansatz for a form with a p-form field strength then 
the dimensionalities of a subset of the submanifolds must sum to $p$.
\item If we have an electic ansatz for a form with a p-form field strength then 
the dimensionalities of a subset of the submanifolds must sum to $p-1$.
\end{itemize}

These conditions will be added to when we come to consider boundary conditions at 
the regular pole of our instantons. The magnetic ansatz automatically obeys the form field equations of motion. In 
the Electric case the equation of motion for E is,

\begin{eqnarray}
E'' + \left( \sum_{i=1}^{T} n_i \frac{a_i'}{a_i} - 2 \sum_{i=\textnormal{lit}} n_i 
\frac{a_i'}{a_i} \right) E' = 0.
\end{eqnarray}

Thus we have,

\begin{eqnarray}
\label{eeqn3}
E' = \frac{C}{ \prod_{i = 1}^T a_i^{n_i} \prod_{j = \textnormal{lit}} 
a_j^{-2 n_j}},
\end{eqnarray}

where C is a constant parameter like M. However, unlike M we will take C here to 
be imaginary. There has been a controversy in the literature over 
the procedure of taking the electric field to be imaginary on Euclideanisation. 
There are several reasons for making this choice. Firstly it means that under 
naive analytical continuation back to Lorentzian signature, treating the continuation as an imaginary general coordinate 
transformation, we will obtain real 
electric fields rather than imaginary ones. Secondly if this choice is not 
made (i.e. if the electric field is taken to be real in Euclidean space) then 
the field strength governed by C in the dominant instanton is infinite - the 
path integral becomes ill defined even when we ignore the conformal factor 
problem in the gravitational sector. When carrying out the calculations (which 
are given below for the case of imaginary electric field) to see this it is 
vital to remember the boundary term for the form field. There are a number of other reasons for this choice of C which 
are less central to the discussion here. By choosing the electric field to be imaginary we make the processes of 
dualisation and Wick rotation commute - this is not the case if we take C 
real. Choosing the electric field to be imaginary seems unavoidable if we want to consider electric/magnetic duality in 
the Euclidean signature 
theory. For example electrically charged extremal black holes can only exist in the 
Euclideanised theory if electric charge, and therefore electric field, is 
imaginary. Magnetically charged extremal black holes however do exist for real 
magnetic charge. A more detailed discussion of these points can be found in refs 
\cite{hawking98,stelle2,stelle1,ross,duncan,hawk,duff2}.

Writing out the remaining equations of motion explicitly including form field 
degrees of freedom we obtain,

\begin{equation}
\label{ab_constraint3}
\frac{1}{2} \left( \sum_{i=1}^T n_i \frac{a'_i}{a_i} \right)^2 - \frac{1}{2} 
\sum_{i=1}^T n_i \left( \frac{a'_i}{a_i} \right)^2 - \sum_{i=1}^T \frac{n_i 
\Lambda_i}{2 a_i^2} = \frac{1}{2} \phi'^2 - {\cal V} - \frac{(M^2 - E'^2)}{2} 
\frac{1}{ \prod_{i= \textnormal{lit}} a_i^{2 n_i}},
\end{equation}
\begin{eqnarray}
\label{aeqn3}
\sum_{i=1}^T n_i \frac{a''_i}{a_i} - \frac{a''_k}{a_k} + \frac{1}{2} \left( 
\sum_{i=1}^T n_i \frac{a'_i}{a_i} \right)^2 - \sum_{i=1}^T n_i \frac{a'_i 
a'_k}{a_i a_k} - \frac{1}{2} \sum_{i=1}^T n_i \left( \frac{a'_i}{a_i} \right)^2 
+ \left( \frac{a'_k}{a_k} \right)^2 - \frac{1}{2} \sum_{i=1}^T \frac{n_i 
\Lambda_i}{a_i^2} + \frac{\Lambda_k}{a_k^2} \\
\nonumber = - \frac{1}{2} \phi'^2 - {\cal V} \pm \frac{(M^2 + E'^2)}{2} ,
\frac{1}{\prod_{i= \textnormal{lit}} a_i^{2 n_i}}
\end{eqnarray}
\begin{equation}
\label{phieqn3}
\phi''+ \sum_{i=1}^T n_i \frac{a_i'}{a_i} \phi' = \frac{\partial {\cal
V}}{\partial \phi}.
\end{equation}

In Eq. (\ref{aeqn3}) the upper sign should be used when the submanifold associated with k is lit and the lower sign when 
it is not. Now using Eq. (\ref{anz1}) and Eq. (\ref{anz2}) we can rewrite the Euclidean action of our system as follows.

\begin{eqnarray}
\label{tot_action3}
S_{\rm{E}}&=&V_{(1)}V_{(2)}...V_{(T)}\left\{
-\left[\frac{\partial\beta}{\partial \xi}\right]_{\xi_S}^{\xi_{N}}
+\int {\rm d}\xi\beta(\xi)\left[-\frac{1}{2}{\cal R}
                     +\frac{1}{2}\phi'^2+{\cal V}(\phi) + \frac{1}{2 d_f!} 
F_{\mu_1...\mu_{d_f}}F^{\mu_1...\mu_{d_f}} \right]
\right\} 
+ \textnormal{ FF Boundary Term}
\end{eqnarray}

\begin{eqnarray}
\label{eff_action3}
= V_{(1)}...V_{(T)}\left\{
-\left[\frac{\partial\beta}{\partial \xi}\right]_{\xi_S}^{\xi_{N}}
-\frac{1}{n_{(1)}+...+n_{(T)}-1}\int {\rm d}\xi \beta(\xi) \left( 2 
{\cal V(\phi)} - \frac{(d_f-1)(M^2 + E'^2)}{\prod_{i=lit} a_i^{2 n_i}}
\right) \right\} + \textnormal{FF Boundary Term}
\end{eqnarray}

where

\begin{equation}
\label{beta3}
\beta(\xi)=a_{(1)}^{n_{(1)}}a_{(2)}^{n_{(2)}}...a_{(T)}^{n_{(T)}}.
\end{equation}

Consider now the form field boundary term.

\begin{eqnarray}
-\int_{\cal \partial M} d^{d-1} x \sqrt{h} 
\frac{1}{d_f!} F^{\mu_1...\mu_{d_f}} n_{\mu_1} A_{\mu_2...\mu_{d_f}} = - \int_{\cal M} d^dx \partial_{\mu_1} \sqrt g 
\left( \frac{1}{d_f!} F^{\mu_1 ... \mu_{d_f}} A_{\mu_2 ... \mu_{d_f}} \right)
\end{eqnarray}

For the cases we are considering the only 'boundary' is in the '0' direction so we may replace this with,

\begin{equation}
\textnormal{FF Boundary term} = - \int_{\cal M} d^dx \partial_{0} \sqrt g \left( \frac{1}{d_f!} F^{0 \mu_2 ... 
\mu_{d_f}} A_{\mu_2 ... \mu_{d_f}} \right).
\end{equation}

For an instanton, $ \partial_{\mu_1} \sqrt g F^{\mu_1...\mu_{d_f}} = 0$ by the form field equations of motion. Thus we 
have,

\begin{equation}
\textnormal{FF Boundary term} = - \int_{\cal M} d^dx \sqrt g \frac{1}{d_f!} F^{0 \mu_2... \mu_{d_f}} \partial_0 
A_{\mu_2... \mu_{d_f}}.
\end{equation}

Now $\partial_0 A_{\mu_2... \mu_{d_f}} = 0$ in the magnetic case so there the boundary term vanishes. In the electric 
case the boundary term becomes,

\begin{equation}
\textnormal{FF Boundary term} = -V_{(1)}V_{(2)}...V_{(T)} \int_{\cal M} d^dx \beta(\xi) \frac{1}{d_f!} 
F_{\mu_1...\mu_{d_f}}F^{\mu_1...\mu_{d_f}}.
\end{equation}
Thus the boundary term reverses the sign of the electric field term in the bulk action \cite{duff2}. Finally then we are 
left with the following expression for the action of our solutions.

\begin{equation}
S_E = V_{(1)}V_{(2)}...V_{(T)}\left\{
-\left[\frac{\partial\beta}{\partial \xi}\right]_{\xi_S}^{\xi_{N}}
-\frac{1}{n_{(1)}+n_{(2)}+...+n_{(T)}-1}\int {\rm d}\xi \beta(\xi) \left( 2 
{\cal V(\phi)} - \frac{(d_f-1)(M^2 - E'^2)}{\prod_{i=lit} a_i^{2 n_i}}
\right) \right\}
\end{equation}

It should be remembered that $E'$ is imaginary in the above.

If we wish to have one regular pole in our instantons, as opposed to having a singularity at each end, then there are 
certain rules which the fields must obey at that regular pole. For the metric and 
scalar field these rules have already been worked out, for the case of higher dimensional examples, and can be found for 
example in \cite{us}. Although instantons with two singular poles are possible \cite{linde} the interesting features we 
wish to discuss are found in the case where only the north pole is singular. The rules are as follows.

\begin{itemize}
\item The Instanton must be ended by the scale factor of one, and only one, of 
the submanifolds going to zero - this submanifold `pinches off' 
the regular pole.
\item The submanifold which pinches off the regular pole must be positively 
curved.
\item The scale factor of the submanifold which pinches off the regular pole 
must approach zero as 
	\begin{eqnarray}
	\label{lambda_constraint}
	a_{(1)}(\xi \rightarrow 0) \rightarrow 
	\sqrt{\frac{\Lambda_{(i)}}{n_i -1}}\;\;\xi,
	\end{eqnarray}
 as we approach the regular pole.
\item All other scale factors must approach a constant at the regular pole.
\item The scalar field must approach a constant at the regular pole.
\end{itemize}

	In the case where form fields are present these rules are supplemented 
by the following derived by demanding regularity and using equations 
(\ref{eeqn3}), (\ref{ab_constraint3}) and (\ref{aeqn3}).
	
\begin{itemize}
\item In the electric case the form must span the submanifold which pinches off 
the regular pole. This means that the form can only span one submanifold (and 
time) and that the submanifold in question must have dimension p-1 
for a p form field strength. In particular this means it is impossible to have 
different dimensional electric field strengths present on the same instanton 
solution.
\item In the magnetic case the form can span any number of submanifolds 
(subject to the rules already outlined above) as long as none of them pinch off the regular 
pole.
\end{itemize}

The equations of motion for the instanton solutions are complicated, so we have solved them numerically. An important 
constraint is that an instanton solution which makes a sensible contribution to the path integral must have finite 
action. Thus we check that as we approach the 
singular pole the fields and scale factors behave in such a way as to render the 
path `integrable'. We did this in \cite{us} for the case of a scalar field in 
the metric in use here. There it was found that, for a wide class of potentials 
for the scalar field, the action was always integrable. \footnote {There are examples where this is not the case. For 
example in the case of higher dimensional generalisations of the Hawking Turok solutions the analysis of Hawking and 
Reall \cite{HR} can be generalised to give the following result. For the action of an instanton with a potential for its 
scalar field of the form $V_0 e^{\alpha \phi}$ to be integrable, we require $|\alpha|<2 \sqrt \frac{d-1}{d-2}$. The same 
result for $d=4$ was also obtained in \cite{saffin98}} This is no longer the case once we include the more general form of 
matter 
being considered here.
For simplicity we consider the case of two submanifolds. Let us assume that near the singularity we have,

\begin{equation}
\label{assymp}
a_1(\xi\rightarrow\xi_N)\propto(\xi-\xi_N)^p\;\;\;\;\;
a_2(\xi\rightarrow\xi_N)\propto(\xi-\xi_N)^q.
\end{equation}

We also make the assumption, which we can check is consistent once we have finished, that in Eq. 
(\ref{ab_constraint3}), Eq. (\ref{aeqn3}) and Eq. (\ref{phieqn3}) the ${\cal V}$ term is negligible near the singularity, 
integrating equation \ref{phieqn3} then gives,

\begin{equation}
\label{phiassymp}
\phi'(\xi\rightarrow\xi_N)\propto(a_1^{n_1}a_2^{n_2})^{-1}.
\end{equation}

We then split the possible cases into three types, $p>0$, $p=0$, and $p<0$. For $p>0$ the terms present due to the form 
fields in the equations of motion and constraint equation are subdominant at the singularity. Consistency of our 
equations then implies $np+mq=1$, from which it follows that the action is integrable. For $p=0$ we again find $np+mq=1$ 
but now the form term in the action is found to be infinite. For $p<0$ we find $mq=1$ and again the form term in the 
action is infinite. Thus when we present results below we only plot the action for configurations satisfying $p>0$. It is 
amusing to note that while the scalar field diverges at the north pole in the cases where the action is finite it 
does not in the case $p<0$ where the action is infinite!

\subsection*{The effect of form fields on the minimum action instanton.}

In the numerical work including form fields, for simplicity, we choose a potential ${\cal V} = \frac{1}{2} \phi^2$ and 
specialise to the case of two submanifolds. Our conclusions are not qualitatively dependent on these choices.
In ref \cite{us} we were able to plot the reduced action of our instantons against the 2 parameters which described 
them, the value of the scalar field and one of the scale factors at the regular pole. Unfortunately such a neat 
representation of our results is not possible here as the parameter space of solutions for a given dimensionality of 
form field, $d_f$, has a maximum dimension of three. The possible parameters are: the value of the non-zero scale factor 
at the regular pole, the value of the scalar field at the same point and M or C if the dimensionalities of the 
submanifolds and field strengths are such as to allow the relevant ansatze.
To find the solution with minimum reduced action in such a parameter space we adopt the following procedure. The reduced 
action for each solution is obtained and the set of parameters which give the instanton of lowest reduced action are 
stored. We then apply a number of checks to make sure that we are obtaining a sensible result. The first of these is 
that we make sure that the minimum action solution is located well within the volume of parameter space which we have 
covered. This ensures that we are indeed looking at a minimum rather than just one of the solutions on the edge of our 
parameter space volume in the direction in which the action is decreasing. Once this is done we can plot a number of 2 
parameter cross-sections through the space of solutions at the minimum to make sure everything is well behaved. Figures 
1 and 2 are representive cross-sections a little way away from the minimum for the case of two 
Einstein metrics of dimensions $n_1 = 3$ and $n_2 = 2$ and a 4 form field strength. This means we are allowed an 
electric but not a magnetic field within the instanton. To be specific we have chosen to take the value of $\Lambda_i = 
n_i -1$, which is the appropriate value for the round metric on $S^{n_i}$. We shall explain the reason for this choice 
and the reason why the cross-sections we have chosen to present are a little way from the minimum shortly.

\begin{figure}[ht]\centering\leavevmode\epsfysize=6cm 
\label{formplot1}
\epsfbox{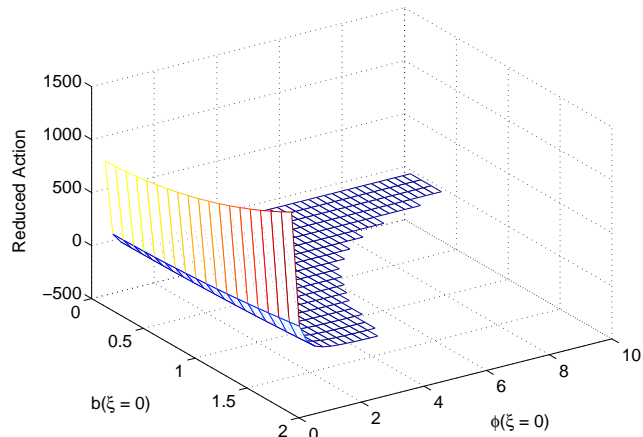}\\
\caption{Plot of the reduced action on a cross-section through parameter space corresponding to C = 0.01i. In this case 
$n_{(a)}= 3, n_{(b)} = 2$ and the antisymmetric field strength has four indices.}
\end{figure}

\begin{figure}[ht]\centering\leavevmode\epsfysize=6cm 
\label{formplot2}
\epsfbox{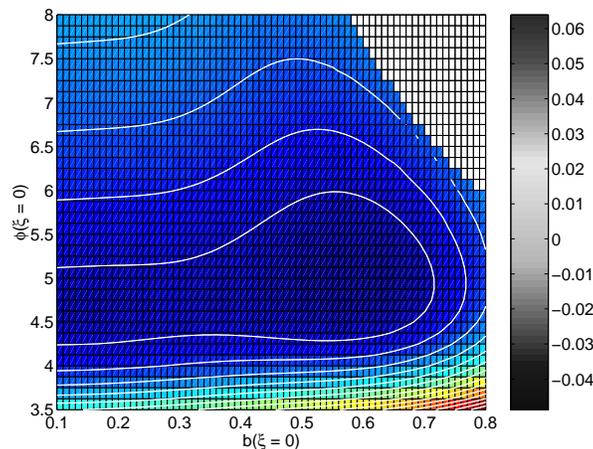}\\
\caption{Magnification of a region of Fig.1 showing the minimum reduced action configuration on this slice of 
parameter space.}
\end{figure}

The second of these plots is a magnification of some structure present on the first, showing a minimum on this slice of 
parameter space with negative reduced action.
The cut out region in the plots corresponds to those configurations which 
do not have finite action.

Now in order to find the dominant configuration in the Hartle Hawking path integral we should, in principle, obtain the 
volume factors of the submanifolds in order to calculate the full action. Infact there is no need to do this here. This 
is because whichever of the metrics considered in this paper we look at, we have found that the dominant instanton, that 
with lowest action, is the one in which C 
and M - all the form fields - vanish. This is the reason why we didn't plot the cross-sections we presented above so as 
to go through the minimum. The plots would then have been identical to those in ref \cite{us}. Given this then we may 
employ the results of \cite{us} for instantons where the only matter is a scalar field. This leads us to conclude that, 
even when including a form field, the dominant instanton of the type studied here is once again a higher dimensional 
generalisation of the simple Hawking Turok solution where the only matter which does not vanish is a scalar field.

\subsection*{The extension of a conjecture.}

In the calculations presented above we have obtained minima of negative reduced action. Thus we can surely obtain 
instantons of arbitrarily negative action merely by using submanifolds of increasingly large volume? We are saved from 
this catastrophe by two things. Firstly we use a result from the theory of Einstein manifolds called Bishop's Theorem. It 
tells us that for a given positive curvature the Einstein metric of largest volume is that of the round sphere. This is 
why we have plotted results appropriate to round spheres above. Secondly we made a conjecture in \cite{us} that {\it 
negative values of the reduced action occur only if $\Lambda_i > 0$ for all i}. We now extend this conjecture to the 
case where forms are present as described in this paper. We have checked this conjecture in numerous cases and have 
always found it to hold true.

\section{Dilatonic couplings to form fields and string effective actions.}

Instead of the case considered above we could allow for couplings between the dilaton and form fields by considering a 
lagrangian density for the antisymmetric tensor fields of 
the form,

\begin{equation}
{\cal L}_FF = f(\phi)F_{\mu_1....\mu_{d_f}}F^{\mu_1....\mu_{d_f}} \frac{\sqrt g}{2 d_f!},
\end{equation}

where $f(\phi)>0$, for all $\phi$ if the Hartle Hawking path integral is to be well defined, even 
if we ignore the conformal factor problem. This condition is satisfied by the low energy effective actions of string 
theory. The derivation of the equations of motion which follow from our new action is a straight forward generalisation 
of the calculation performed in the previous two sections. We obtain the following:

\begin{eqnarray}	
\label{eeqn2}
E' = \frac{C}{ \prod_{i = 1}^T a_i^{n_i} \prod_{j = \textnormal{lit}} 
a_j^{-2 n_j} f(\phi)}
\end{eqnarray}

\begin{equation}
\label{ab_constraint2}
\frac{1}{2} \left( \sum_{i=1}^T n_i \frac{a'_i}{a_i} \right)^2 - \frac{1}{2} 
\sum_{i=1}^T n_i \left( \frac{a'_i}{a_i} \right)^2 - \sum_{i=1}^T \frac{n_i 
\Lambda_i}{2 a_i^2} = \frac{1}{2} \phi'^2 - {\cal V} - \frac{(M^2 - E'^2)}{2} 
\frac{f(\phi)}{ \prod_{i= \textnormal{lit}} a_i^{2 n_i}}
\end{equation}

\begin{eqnarray}
\label{aeqn2}
\sum_{i=1}^T n_i \frac{a''_i}{a_i} - \frac{a''_k}{a_k} + \frac{1}{2} \left( 
\sum_{i=1}^T n_i \frac{a'_i}{a_i} \right)^2 - \sum_{i=1}^T n_i \frac{a'_i 
a'_k}{a_i a_k} - \frac{1}{2} \sum_{i=1}^T n_i \left( \frac{a'_i}{a_i} \right)^2 
+ \left( \frac{a'_k}{a_k} \right)^2 - \frac{1}{2} \sum_{i=1}^T \frac{n_i 
\Lambda_i}{a_i^2} + \frac{\Lambda_k}{a_k^2} \\
\nonumber = - \frac{1}{2} \phi'^2 - {\cal V} \pm \frac{(M^2 + E'^2)}{2} 
\frac{f(\phi)}{\prod_{i= \textnormal{lit}} a_i^{2 n_i}}
\end{eqnarray}

\begin{equation}
\label{phieqn2}
\phi''+ \sum_{i=1}^T n_i \frac{a_i'}{a_i} \phi' = \frac{\partial {\cal V}}{\partial \phi} + \frac{(M^2 + E'^2)}{2} 
\frac{1}{\prod_{i= \textnormal{lit}} a_i^{2 n_i} }\frac{\partial f}{\partial \phi}
\end{equation}

The action may be written as,

\begin{equation}
S_E = V_{(1)}V_{(2)}...V_{(T)}\left\{
-\left[\frac{\partial\beta}{\partial \xi}\right]_{\xi_S}^{\xi_{N}}
-\frac{1}{n_{(1)}+n_{(2)}+...+n_{(T)}-1}\int {\rm d}\xi \beta(\xi) \left(2 {\cal V} - \frac{(d_f-1)(M^2 - 
E'^2)f(\phi)}{\prod_{i=lit} a_i^{2 n_i}}
\right) \right\}.
\end{equation}

It should be remembered that $E'$ is imaginary in the above. As before, we first of all check which of our configurations 
have finite action. Using Eq. (\ref{assymp}) and assuming $f \propto (\xi - \xi_n)^r$ a similar analysis to that presented 
earlier shows that for the action to be finite we require $2np>r$. If this condition is fullfilled then $np+mq=1$. 
$\phi'$ obeys Eq. (\ref{phiassymp}) and so assuming that the condition holds and knowing f as a function of $\phi$ we can 
calculate r. We can then check for consistency by looking to see that we do indeed have $r<2np$, which means that 
integrable solutions could exist.

In the case where ${\cal V}$ is non-zero and gives rise to integrable instantons (${\cal V} = \frac{1}{2} \phi^2$ for 
example) we once again find that the form fields vanish in the dominant instanton.
A more interesting case is that where 
we have ${\cal V} = 0$ and the potential for the scalar field is completely generated by its coupling to the 
antisymmetric tensor fields. The reason for this is twofold. Firstly such a situation arises if we naively try and find 
instantons which extremise Euclidean low energy string effective actions. Secondly it appears at first glance that such 
a situation should give us dominant instantons with constant $\xi$ cross-sections of non-trivial topology. This is because 
compact singular instantons can only exist, in the absence of branes, when $\phi$ has a potential. If 
$\phi$ is to have a potential here then the form fields must be non-zero. This in turn leads to the instantons 
necessarily having non-trivial 'spatial' topology unless the rank of the form field strength is the same as, or one less 
than, the dimensionality of spacetime.

	Unfortunately once we start to look at the solutions things do not work out quite how we might hope. The problem 
that arises is that there are no instantons of the type being described here which have finite action. 

We numerically integrated our equations as before and checked wide ranges of parameter space looking for integrable 
instantons, failing to find any for the choices of the function $f(\phi)$ that we tried. For example $f(\phi) = 0.5 
\phi^2$ or any of the exponential couplings appropriate for a 10 dimensional low energy effective action from string 
theory. The only types of field profile we obtain are non-integrable ones or configurations like those shown in figures 
3 and 4.

\begin{figure}[ht]\centering\leavevmode\epsfysize=5.5cm 
\label{formplot4}
\epsfbox{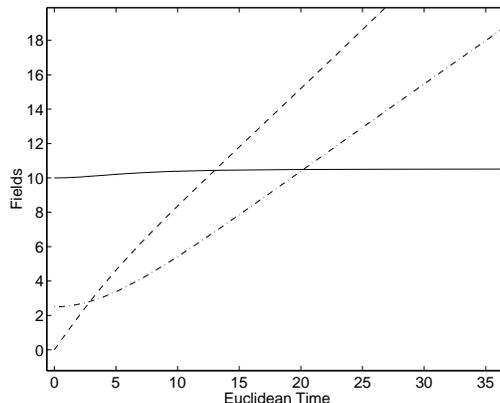}\\
\caption{Field profiles for one of the non-compact instantons. The solid line is $10 \phi$ and the dashed and dot-dashed 
lines are scale factors of 3 and 2 dimensional spheres respectively. In this case we have $c=i$ and $f = e^{-0.5 
\phi}$.}
\end{figure}

\begin{figure}[ht]\centering\leavevmode\epsfysize=5.5cm 
\label{formplot3}
\epsfbox{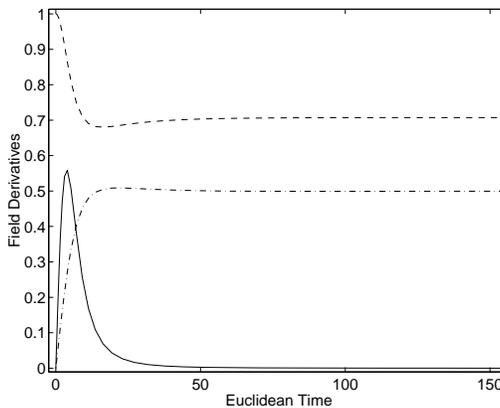}\\
\caption{Curvature driven terminal velocities of the fields shown in the previous plot. The solid line is the derivative 
of the solid line in fig.3 and so on.}
\end{figure}

The explanation of what is happening is as follows. If the instanton does not close off quickly enough then the scale 
factor b becoming large 'turns off' the potential of the scalar field due to the presence of the form fields. $\phi$ 
then skids to a halt and there is no energy to close off the instanton. The two scale factors head for curvature driven 
terminal velocity and we can see that this is the asymptotic form of the solution. In fact it is easy to predict the 
late time rate of change of each scale factor. If we put double derivatives and the form and $\phi'$ terms to zero in 
our equations and rearrange slightly we obtain the following.

\begin{eqnarray}
(m-1)\left( \frac{b'}{b} \right)^2 + n \frac{a'b'}{ab} - \frac{\Lambda_m}{b^2} = 0 \\
(n-1)\left( \frac{a'}{a} \right)^2 + m \frac{a'b'}{ab} - \frac{\Lambda_n}{a^2} = 0
\end{eqnarray}

If we now make the assumption $a=kb$ where k is a constant (which will be true at late times if the system approaches 
terminal velocity) we obtain,

\begin{eqnarray}
\label{termvel}
a' = \sqrt \frac{\Lambda_n}{D-2}   \;\;\;\;\;\;\; b' = \sqrt \frac{\Lambda_m}{D-2}.
\end{eqnarray}

For the parameters appropriate for the results plotted in figures 3 and 4 we obtain $a' = \frac{1}{\sqrt 2}$ and $b = 
\frac{1}{2}$. We see that this is in agreement with the numerical results. It should be noted that Eq. (\ref{termvel}) 
seems to imply that we wouldn't see this behaviour for negatively curved submanifolds. This is indeed the case, for such 
metrics, and those where some of the submanifolds have zero curvature, we find only compact non-integrable 
configurations.

	The low energy effective actions of the superstring in the Einstein frame are special cases of this type of 
system, if we keep only the kinetic terms for the form fields \cite{low}. We have not been able to find any integrable 
singular instantons of the type described here at all for these systems. Naively, we might conclude that we can not use 
the instanton approximation to the Hartle Hawking wavefunction in string theory to verify the existence of hidden extra 
dimensions. However, there are a number of possible resolutions that have not yet been explored. These include the need to 
include corrections to the low energy effective action as we approach the singular pole. Unless these corrections stop the 
curvature from reaching 
the string scale we would have to include an infinite number of them - our effective field theory would 
break down altogether.

\section{conclusions.}

	We have introduced antisymmetric tensor field degrees of freedom into the metric considered in \cite{us}. In the 
case of a simple $F^2 \sqrt g$ lagrangian term for these fields we find that they vanish in the dominant instanton. We 
have also studied the case where there is some kind of dilatonic coupling between the scalar field and the forms. In the 
case where the scalar fields potential is completely generated by its coupling to the forms we find no finite action 
singular instantons at all. Special cases of such systems include the naive string low energy effective actions.
	The main conclusion to be drawn from our calculations is that a naive application of the Hartle Hawking proposal 
to these higher dimensional theories does not appear to lead to a universe with compactified extra dimensions as we 
would like. The dominant instanton in the path integral seems to have such a high degree of symmetry that it continues 
to a universe where none of the extra dimensions are compactified. It would be interesting to investigate the impact 
higher order corrections would have when applying the no boundary proposal in higher dimensions.

\section*{Acknowledgements}

We would like to thank Paul Saffin for some useful comments. J.A.G. is supported by PPARC. The numerical work was carried 
out on the SGI Origin platform using COSMOS Consortium 
facilities, funded by HEFCE, PPARC and SGI. We acknowledge computing support from the Sussex High Performance Computing 
Initiative.

\end{document}